# Single-crystal elastic moduli, anisotropy and the B1-B2 phase transition of NaCl at high pressures: Experiment vs. *ab-initio* calculations

Feng Xu,[1,]* Laurent Belliard,[2] Chenhui Li,[3,4] Philippe Djemia,[3] Loïc Becerra,[2] Haijun Huang,[1] Bernard Perrin[2] and Andreas Zerr[3,]*

[1]School of Science, Wuhan University of Technology, 430070 Wuhan, China

[2]Institut des NanoSciences de Paris, CNRS UMR 7588, Sorbonne Université, 75005 Paris, France

[3]Laboratoire des Sciences des Procédés et des Matériaux, CNRS UPR 3407, Université Sorbonne Paris Nord, Alliance Sorbonne-Paris-Cité, 93430 Villetaneuse, France

[4]School of Materials Science and Engineering, Beijing Institute of Technology, 100081 Beijing, China

**Abstract**

Single-crystal elastic moduli, $C_{ij}$, and the B1-B2 phase transition of NaCl were investigated experimentally, using time-domain Brillouin scattering (TDBS), and theoretically, via density-functional-theory (DFT), to 41 GPa. Thus, we largely extended pressure range where $C_{ij}$ and elastic anisotropy of the solid are measured, including the first experimental data for the high-pressure B2 phase, NaCl-B2. NaCl-B1 exhibits a strong and growing with pressure anisotropy, in contrast to NaCl-B2. Theoretical values obtained using different advanced DFT functionals were compared with our measurements but no one could satisfactorily reproduce our experimental data for NaCl-B1 and NaCl-B2 simultaneously. For all available DFT results on the principal shear moduli and anisotropy, the deviation became pronounced when the degree of compression increased significantly. Similar deviations could be also recognized for other cubic solids having the same B1-type structure and similar bonding, such as CaO, MgO, or $(Mg_{1-x},Fe_x)O$. Furthermore, the available experimental data suggest that the B1-B2 phase transition of NaCl and the above mentioned compounds are governed by the Born stability criterion $C_{44}(P) - P > 0$.

## I.   INTRODUCTION

NaCl, being the archetype of ionic solids, is the material of choice for investigating specifics of properties and phase transitions of multiple alkali halides, alkaline-earth oxides (MgO, CaO etc.), and transition metal nitrides/carbides (TiN, TiC etc.) crystallizing in the cubic B1 structure. Detailed examinations of high-pressure behavior of this diversity of compounds permits extraction and evaluation of contribution of composition, bonding, and structure to physical properties (e.g. elasticity, conductivity, hardness) of solids in general, and NaCl in particular. NaCl is also remarkable because it is the most abundant nonoxide compound on the Earth's surface, if that dissolved in oceans is included [1,2], and because equation of state (EOS) of NaCl-B1 calculated by Decker 50 years ago [3] is recommended as the primary pressure standard to $P < 30$ GPa [4]. Above this pressure, NaCl-B1 transforms to NaCl-B2 having CsCl-type structure [5], also proposed as a pressure standard [6]. Even though the Decker's EOS and other pressure scales (e.g. ruby fluorescence scale [7]) are known to be of insufficient accuracy, none of the modern EOSes obtained using various DFT approaches was proposed as a better substitute. Alkali halides are considered as problematic and search for appropriate DFT functionals to describe such ionic and relatively weakly bonded systems remains a subject of theoretical efforts [8,9]. Presently, experimental values obtained at atmospheric or low pressures are predominantly used for evaluation of various DFT approaches e.g.[9-11] even though strong compression provides deeper insights into bonding in any solid, including NaCl. However, just two experimental EOSes of NaCl-B1, reported more than 40 years ago, are available: One, based on direct measurements of pressure and volume, is limited to 3.2 GPa [12]. The other was derived from shock compression to 25 GPa with experimental uncertainty of $\geq$10% and bulk modulus, $B$, ~7% below the Decker's value (see [13] and Fig. S2 in Supplemental Material).

We show that a much stronger experimental criterion of reliability of any DFT approach is pressure dependence of the principal shear moduli which are, in case of a cubic solid, $C_{44}(P)$ and $C'(P)$, where $C' = (C_{11} - C_{12})/2$. NaCl also enables application of the second reliability criterion, namely, whether one and the same DFT-approach predicts reasonable $C_{44}(P)$ and $C'(P)$ for two phases of one and the same compound. In the case of NaCl, these are the atmospheric-pressure B1 phase and the high-pressure B2 phase. The criteria could not be applied earlier because pressure dependences of single crystal elastic moduli $C_{ij}(P)$ of NaCl-B1 (and, accordingly, of $C_{44}(P)$ and $C'(P)$) were limited to $P = 17$ GPa [14-16] (see below) where the interatomic distances decrease by <10% only and the pressure effect is comparable with experimental uncertainties. Moreover, reliable experimental $C_{ij}(P)$ of NaCl-B2 were not available and, to our knowledge, no experimental $C_{ij}(P)$ of the B2 phase of any other compound have been reported yet. Besides, scarce reports on average sound velocities in polycrystalline NaCl-B1, $V_{L(avg)}$, [6,17-19] are inconclusive because the growing with pressure elastic anisotropy, quantified using the Zener ratio $A = C_{44}/C'$, induces systematic errors [20].

Most of high-pressure investigations of NaCl applied X-ray diffraction (XRD) to examine structure evolution and the B1-B2 phase transition [5,21-24]. But the transition mechanism was not proposed even though the same mechanism is expected to govern the B1-B2 transitions in other compounds such as CaO and MgO. The latter compound is of outstanding importance for geosciences because it is the end member of the solid solution $(Mg_{1-x},Fe_x)O$, called ferropericlase, one of the two most abundant nonconducting compounds of our planet [25].

The straight-forward way to access $C_{ij}(P)$, and thus $C_{44}(P)$ and $C'(P)$, is measurement of longitudinal- and transversal sound velocities, $V_L$ and $V_T$, respectively, along different crystallographic directions in a compressed single crystal [26]. Ultrasonic techniques provide such $V_L(P)$ but require big samples limiting the measurements to $P < 10$ GPa [27]. Spontaneous Brillouin light scattering (BLS) and impulsive-stimulated scattering are other methods to measure $C_{ij}(P)$ of transparent solids in a broad $P$-range but only if single crystals persist on compression in a diamond anvil cell (DAC) [28,29]. However, even when single crystals persist upon compression they inevitably degrade to polycrystals upon a phase transition [30,31]. In such case, a peak in a classical BLS spectrum represents (due to a limited resolution along the beam path) a sum of contributions from differently oriented crystallites. Inelastic X-ray scattering (IXS) is another technique which permits measurement of $V_L$ of transparent solid samples at high pressures in a DAC but it also necessities single-crystalline samples to access $C_{ij}(P)$ [32]. In contrast, time-domain Brillouin scattering (TDBS), based on a pump-probe method [33], permits measurement of $V_L(P)$ and $V_T(P)$ of transparent polycrystalline samples in a DAC with a high axial resolution and thus to access $C_{ij}(P)$ applying the envelope method [34-38]. Here, we used this technique to measure $C_{44}(P)$ and $C'(P)$ of NaCl-B1 and NaCl-B2 to 41 GPa, the pressure region previously not accessible for measurements of these physical parameters. Our experiments were accompanied by *ab-initio* calculations applying various DFT approaches, with and without van-der-Waals (vdW) forces which contribution was reported to be significant in compressed molecular solids, e.g. $H_2O$ ice [39].

## II. METHODS

### A. Samples and TDBS technique

In this work, high pressures were generated using a DAC with beveled diamond anvils (culet of 300 μm in diameter). A hole of ~100 μm in diameter drilled in a 40 μm-thick pre-indented gasket served as the sample volume. The volume was filled with NaCl grains without any additional pressure transmitting medium (PTM) (see inset of Fig. 1) and an absorbing opto-acoustic transducer was placed on one of the diamond anvils, in contact with the polycrystalline NaCl sample. We used, as transducers, thin metallic films which have been sputtered on a 10-μm-thick single-crystal Si substrate or directly on one of the diamond anvils. Ruby grains used to measure pressure were placed around the transducer, close to locations of the TDBS measurement. This limited uncertainties caused by pressure gradients: According to recent reports, yield stress of NaCl is, within the pressure range of our experiments and experimental uncertainties, the same as yield stresses of the softest known PTMs, such as solid helium, neon or argon [40]. Thus, use of an additional PTM was dispensable in our measurements. Applying a model of pressure gradients in a sample compressed in a DAC, as described in a classical book [41], we calculated the upper limit of pressure differences $P_{diff}$ between locations of our TDBS measurements and the ruby grains used for the pressure measurements: $P_{diff} = \sigma_y d/h$, where $d \leq 30$ μm is the maximal distance between locations of the TDBS measurements and the ruby grains, $h \geq 14$ μm is thickness of the whole sample (thickness of the NaCl layer was ~7 μm at least while that of the transducers ~7 μm [30]. Using $\sigma_y \sim 0.5$ GPa, yield stress of the sample material at the maximal pressure of our experiments of $P = 41$ GPa [40], we obtained $P_{diff} \leq 1.1$ GPa or $\leq 2.7$ % of the pressure value.

In our TDBS measurements, ultrashort acoustic pulses were generated and detected by picosecond laser pulses. In particular, radiation of a mode-locked pulsed Ti:sapphire laser with the wavelength $\lambda = 800$ nm (MAI-TAI Spectra, repetition rate 80 MHz, pulse width 100 fs) was split into a pump- and probe beam. The pump beam was modulated at 1.8 MHz by an acousto-optic modulator and doubled in frequency by a BBO crystal, while the probe beam was delayed by a mechanical translation stage. Both pump and probe beams were focused to spots of ~2 μm in diameter using an objective with a long working distance (magnification ×50) to adapt geometry of the used DAC. The beams passed through the transparent NaCl sample which back-side was in contact with a transducer. An ultrashort pump-laser pulse generated a coherent acoustic pulse (CAP) near the transducer surface which propagated through the NaCl sample. Probe-laser pulses were used to measure the CAP velocity: In particular, we measured transient reflectivity of the sample (in which the CAP was moving) as a function of the time delay with respect to the pump-laser pulse. Brillouin oscillations, as a function of the delay time, resulting from interference of the probe-laser beam scattered by the CAP and by stationary interfaces (see inset of Fig. 1) were measured, at the pump modulation frequency, by a photodetector. The oscillations exhibited frequency $f_B$ proportional to $V_L$ of the sample material at the particular pressure [33]:

$$f_B = 2nV_L/\lambda \qquad (1)$$

where $\lambda$ is wavelength of the probe beam in vacuum (800 nm), and $n$ refractive index of NaCl at this wavelength and pressure. Frequencies corresponding to $V_T$ were not recognized in the collected TDBS signals. A CAP moving in a polycrystalline NaCl sample provided local information on $V_L$ along those crystallographic directions in traversed grains which were parallel to direction of the CAP propagation. Temporal window $\Delta t \approx 1$ oscillation, selected for the signal-frequency evaluation using the Short-Time Fourier Transform (STFT), resulted in axial/depth resolution of ~250 nm which didn't degrade with pressure. Consideration of oscillations with amplitudes exceeding ~35% of the maximal one permitted STFT-analysis of 7-23 depth segments providing independent $f_B$ values, and, accordingly, local $V_L$. It could be argued that nm-sized grains existed or formed in the sample upon compression and influenced the local $V_L$. As was shown in earlier TDBS-experiments on $H_2O$-ice compressed in a DAC to similar pressures [34], crystallites of relatively soft solids (e.g. $H_2O$ or NaCl) compressed in a DAC have typical sizes similar or below ~450 nm. This followed from the observation of complete X-ray diffraction rings collected for a compressed powder sample of cubic $H_2O$-ice and NaCl [40] even though some texture can develop upon compression. On the other hand, because XRD peaks of $H_2O$ and NaCl remain relatively sharp on compression, grain sizes of $H_2O$ or NaCl do not decrease significantly below ~100 nm. Therefore, size of grains in our NaCl sample, between ~100 and ~450 nm, was comparable with axial resolution of our signal processing of ~250 nm. Such axial resolution permitted a reliable extraction of the maximal and minimal $V_L$ values in a single crystal of the examined solid using the envelope method if number of unique $V_L$ measurements was sufficiently large.

For each sample and at each pressure step, we collected one/two TDBS signals, performed their STFT analysis and extracted maximal and minimal frequencies, $f_{B(max)}$ and $f_{B(min)}$, respectively (Fig. 1). Significant scattering within the groups of $f_{B(max)}$ and $f_{B(min)}$ indicated that only a part of the TDBS signals contained depth segments corresponding to crystallites oriented with the fastest (and the highest frequency, $f_{max}$) or the slowest (and the lowest frequency, $f_{min}$) crystallographic directions along the CAP paths. In the present work, six pressure runs provided a large number of independent $f_{B(max)}$ and $f_{B(min)}$ values permitting recovery of $f_{max}$ and $f_{min}$ with a high degree of confidence. A strong deviation of $f_{max}$ from $f_{min}$ in NaCl-B1 (Fig. 1) indicated a pronounced elastic anisotropy when compared with NaCl-B2. Our DFT-calculations showed that velocities along the crystallographic directions $\langle 100 \rangle$ and $\langle 111 \rangle$, $V_{L\langle 100 \rangle}(P)$ and $V_{L\langle 111 \rangle}(P)$, are the fastest and the slowest, respectively, for both phases. Accordingly, their experimental values were derived using Eq. 1 at $P \leq 28$ GPa for NaCl-B1, and at $P \geq 30$ GPa for NaCl-B2. The intermediate pressures, where both phases could coexist, were omitted. For each phase, a second-order polynomial was fitted to the $f_{max}$ data-points to obtain a smooth dependence $f_{max}(P)$ and then $V_{L\langle 100 \rangle}(P)$ applying our theoretical $n(P)$ (Fig. 2, insert). Similarly, a smooth $V_{L\langle 111 \rangle}(P)$ dependence was derived from the $f_{min}$ data-points.

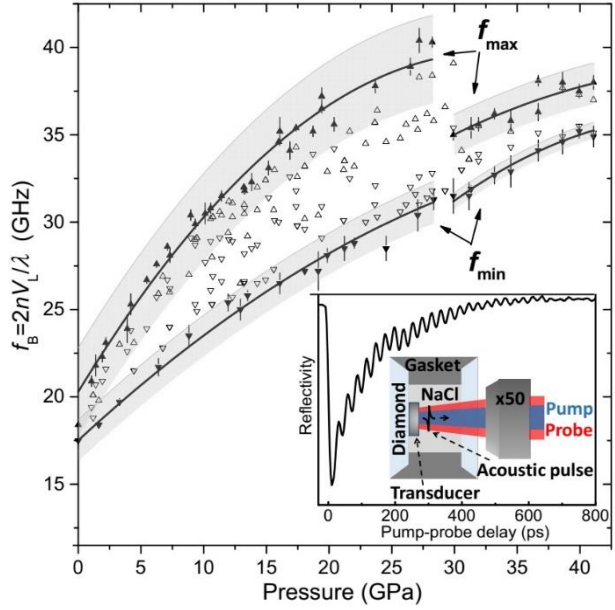

**FIG. 1** Frequencies $f_{B(max)}$ and $f_{B(min)}$, for each TDBS signal collected from the compressed polycrystalline NaCl samples, are represented by triangles showing up and down, respectively. The extremes, $f_{max}$ and $f_{min}$, are highlighted as solid symbols with error bars. Fits of quadratic polynomials, $f_{max}(P)$ and $f_{min}(P)$, to the latter data-points are shown by solid lines. Gray bands represent experimental uncertainties. Inset: A raw TDBS signal showing transient reflectivity of the sample as a function of the pump-probe delay time recorded at $P = 41$ GPa, and schematics of the sample inside a DAC, showing the pump and probe beams.

## B. DFT-calculations

Because pressure dependence of refractive index of NaCl, $n(P)$, is not well established experimentally, we derived it from DFT calculations using the hybrid functional GGA-HSE06 known to provide accurate optical properties [42]. A detailed description of the dielectric-property simulations is given elsewhere [43]. We also calculated relaxed lattice parameters, specific volume, formation enthalpy, $C_{ij}(P)$ of both NaCl-phases to 50 GPa using the Vienna Ab-initio Simulation Package (VASP) [44] (see Supplemental Material for details). We applied two different PBE functionals (GGA-PBE, GGA-PBESol) [45,46], those with various vdW corrections (GGA-PBE-TS-HI [47], GGA-PBE-D2, GGA-PBESol-D2 [48], GGA-PBESol-D3 [49]), and the non-local vdW density functional (OptB86b) [50]. A comprehensive overview of different DFT-vdW methods is given elsewhere [51]. $C_{ij}(P)$ were calculated by the stress-strain method applying moderate strains of 0.001-0.004 and strain matrixes proposed elsewhere [52]. To verify these calculations, we applied the strain-energy method [52] with much larger strains [-0.08, 0.08]. Finally, third-order-elastic constants $C_{ijk}$ were calculated and used in a simplified hydrostatic approach requiring only values at a reference pressure $P_0$, and the derivatives $(dC_{ij}/dP)_{P=P0}$ (see Supplemental Material). In our calculations, both NaCl-B1 and NaCl-B2 were modelled with their primitive unit cells. We fixed the energy cut-off to 520 eV for the plane wave basis set and the electronic energy convergence criteria to $10^{-6}$ eV. Energy convergence tests for $k$-meshes ($10\times10\times10$ and $22\times22\times22$) were performed for all functionals and very large $k$-meshes ($30\times30\times30$ and $40\times40\times40$) for the PBE functionals. The $\Gamma$-centred $30\times30\times30$ $k$-mesh was finally chosen.

## III. RESULTS

### A. Sound velocities

At $P < 8$ GPa, our experimental $V_{L\langle 100 \rangle}(P)$ and $V_{L\langle 111 \rangle}(P)$ of NaCl-B1 (Fig. 2) agreed with previous measurements on single crystals [14-16]. As already recognized for other cubic solids [37,38], the earlier measured $V_{L(avg)}(P)$ of polycrystalline NaCl [17-19] approached $V_{L\langle 111 \rangle}(P)$ because in any cubic crystal (i) directions close to $\langle 110 \rangle$ have the highest multiplicity and thus dominate signal from a polycrystalline sample collected using conventional techniques such as BLS or ultrasonics, (ii) $V_{L\langle 111 \rangle}$ is much closer to $V_{L\langle 111 \rangle}$ than to $V_{L\langle 100 \rangle}$ and the collected BLS peaks are asymmetric, and (iii) the asymmetry becomes pronounced when elastic anisotropy increases. A detailed analysis of the effect of elastic anisotropy on $V_L$-distribution in cubic solids, asymmetry of the BLS peaks and/or shift of the $V_{L(avg)}$ in ultrasonic measurements is given elsewhere [20]. Our measurements could not be affected by sample texturing because it was reported to be weak up to the B1-B2 phase transition [40]. Accordingly, the number of crystallites oriented with $\langle 111 \rangle$ and $\langle 100 \rangle$ along the DAC axis was significant and a possible weak texture effect was compensated by the numerous independent spatially resolved $V_L$ measurements. This is further supported by a good agreement of our data with the unique BLS data-point obtained for a single crystal of NaCl-B1 at $P = 26$ GPa [53] (Fig. 2). The latter fact and the finding that our experimental $V_{L\langle 111 \rangle}(P)$ (the lowest possible $V_L$ in a single crystal of NaCl-B1) exceeds the theoretical ones, exclude the presence of significant amounts of nm-sized grains in our samples. In an opposite case, intercrystalline material (e.g. grain boundaries) strongly reduces, according to earlier measurement on nanocrystalline MgO [54], sound velocities of the entire sample when compared with those of a powder sample composed of bigger grains. The above mentioned IXS measurement provided $V_{L\langle 100 \rangle}(P)$ consistent with our results, but the $V_{L\langle 111 \rangle}(P)$ dramatically deviated, upon compression, both from our and the earlier reported single-crystal BLS data-point at $P = 26$ GPa.

Our DFT-calculations based on the above listed functionals (except GGA-PBE-D2) and the previous ones [40,55] provided $V_{L\langle 100 \rangle}(P)$ of NaCl-B1 agreeing with our measurements within experimental uncertainties. However, most of the calculated $V_{L\langle 111 \rangle}(P)$ are significantly lower than our experimental values at $P > 15$ GPa or even earlier (Fig. 2). Application of the GGA-PBE-TS-HI functional resulted in the strongest deviation of the calculated and experimental $V_{L\langle 111 \rangle}$. The GGA-PBE-D2 functional provided higher $V_L$-values for NaCl-B1: $V_{L\langle 100 \rangle}(P)$ values were slightly higher but the $V_{L\langle 111 \rangle}(P)$ agreed reasonably well with our data. For NaCl-B2, the character of deviation of the DFT-calculations from our measurements was, except three DFT functionals, inverse (Fig. 2): The predicted $V_{L\langle 111 \rangle}(P)$ values were not far from the measured ones but the calculated $V_{L\langle 100 \rangle}(P)$ dependences were significantly overestimated. The GGA-PBESol-D2 functional provided moderately higher values for both $V_{L\langle 111 \rangle}(P)$ and $V_{L\langle 100 \rangle}(P)$. The GGA-PBE-D2 provided much higher $V_{L\langle 111 \rangle}(P)$ and $V_{L\langle 100 \rangle}(P)$ values, in contrast to a reasonable agreement with our measurements on NaCl-B1. The GGA-PBE-TS-HI functional provided consistent results for both NaCl-phases: The predicted $V_{L\langle 100 \rangle}(P)$ was close to our measurements while the calculated $V_{L\langle 111 \rangle}(P)$ deviated strongly (Fig. 2). Summarizing, none of the considered here or earlier used functionals, involving or not vdW forces, could predict reasonably well the dependences $V_{L\langle 111 \rangle}(P)$ and $V_{L\langle 100 \rangle}(P)$ for both NaCl phases simultaneously.

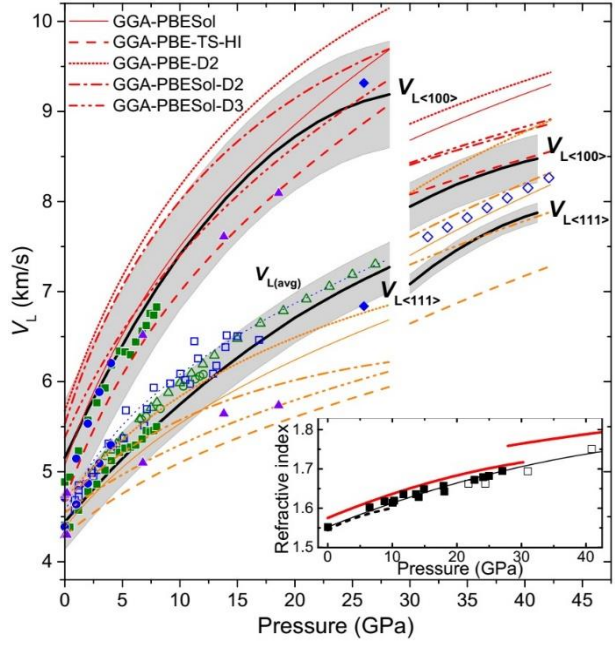

**FIG. 2.** $V_{L\langle100\rangle}$ and $V_{L\langle111\rangle}$ of compressed NaCl-B1 and NaCl-B2: Solid black lines indicate our experimental results and grey bands their uncertainties. They agree with the earlier measurements on single crystals performed using BLS [14,53] (blue solid symbols) and ultrasonics [15] (green solid squares). Violet solid triangles show single-crystal data obtained using IXS [16]. $V_{L\langle100\rangle}(P)$ calculated here using the GGA-PBESol-, GGA-PBE-TS-HI-, GGA-PBE-D2-, GGA-PBESol-D2- and GGA-PBESol-D3 functionals are shown by red solid, dashed, dotted, dash-dotted and dash-dot-dotted lines, respectively, while the calculated $V_{L\langle111\rangle}(P)$ are shown in the same style by orange lines. The results for the GGA-PBESol functional represent also our very similar results obtained using the GGA-PBE- and OptB86b functionals, and the earlier calculations [40,55] (see Fig. S3 in Supplemental Material for more details). Earlier experimental and theoretical $V_{L\langle avg\rangle}(P)$ are shown as follows: ultrasonics - green open triangles [17] and circles [18]; BLS - blue open squares [19] and diamonds (for NaCl-B2) [6]; Eulerian calculations - blue dotted line [56]. Inset: $n(P)$ of NaCl calculated using the GGA-HSE06 functional (red solid lines), compared with earlier shock-compression data: solid squares [57] and open squares [58]. Black dashed line shows extrapolated optical measurements [59], black solid line - theoretical dependence [60].

## B. Elastic moduli and anisotropy

Our experimental $V_{L\langle100\rangle}(P)$ and $V_{L\langle111\rangle}(P)$ combined with $B(P) = (C_{11}(P)+2C_{12}(P))/3$ permitted determining $C_{ij}(P)$, and thus the principle shear moduli $C_{44}(P)$ and $C'(P)$, of both NaCl phases [20]. Here, $B(P)$ of NaCl-B1 and of NaCl-B2 were taken from the Decker's EOS [3] and Ref. [61], respectively. The derived here experimental $C_{44}(P)$ and $C'(P)$ of NaCl-B1 agreed with the earlier low-pressure measurements [14,15], the BLS data-point at $P = 26$ GPa [53] and unpublished data in a thesis work reporting results from radial-XRD measurements on polycrystalline NaCl compressed in a DAC [62]. The IXS technique provided $C'(P)$ agreeing with our observations but the $C_{44}(P)$ deviated significantly with increasing compression. The $C'(P)$ of NaCl-B1, calculated using the GGA-PBESol-, GGA-PBE-, OptB86b-, GGA-PBE-TS-HI- and GGA-PBESol-D3 functionals, matched experiments below ~25 GPa (Fig. 3a). Flattening of our experimental $C'(P)$ at $P > 25$ GPa cannot be explained by a sample-state change because the experimental $C_{44}(P)$ increased monotonically. All considered DFT functionals predicted a negative slope of $C_{44}(P)$ contradicting our experiment, the earlier single-crystal BLS data-point at 26 GPa and the radial-XRD results. This tendency resulted, accordingly, in much lower theoretical $A(P)$ values when compared to our measurement (Fig. 3c). For NaCl-B2, $C'(P)$ was measured to drop significantly while $C_{44}(P)$ changed weakly. All considered DFT functionals provided gravely overestimated $C'(P)$, opposite to NaCl-B1, while only the GGA-PBESol-, GGA-PBE-, OptB86b-, and GGA-PBE-TS-HI functionals predicted tolerable $C_{44}(P)$, comparable with our measurements. Summarising, our and earlier experimental and theoretical results indicated that the existing DFT methods (including or not vdW interaction) provide smaller anisotropy ratios $A = C_{44}/C'$ (Fig. 3c) when NaCl is strongly compressed: In the case of NaCl-B1, the statement is valid for all considered DFT functionals and the difference becomes pronounced upon compression. In the case of NaCl-B2, the statement is not valid for the GGA-PBE-D2-, GGA-PBESol-D2- and GGA-PBESol-D3 functionals but just because they provided much higher $C_{44}(P)$ and thus artificially elevated $A(P)$.

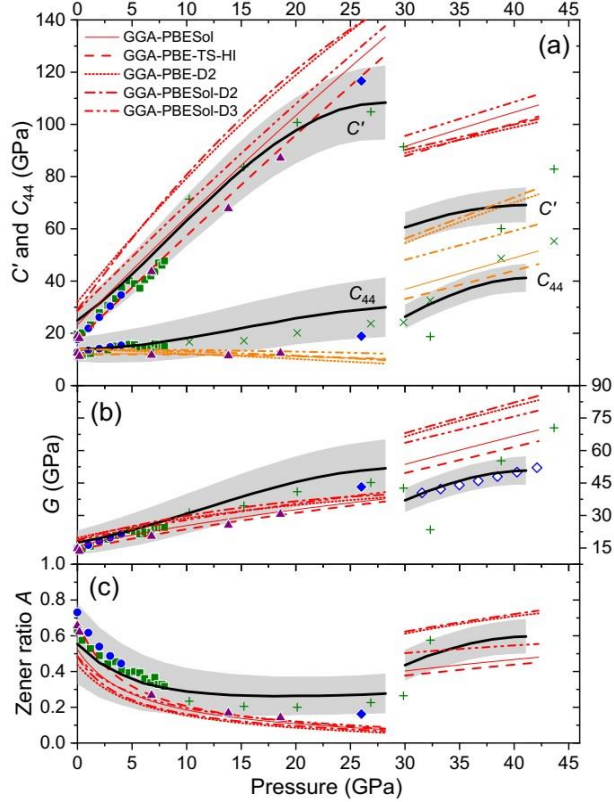

**FIG. 3**. Our experimental and theoretical $C'(P)$, $C_{44}(P)$, $G(P)$ and $A(P)$ of NaCl-B1 and NaCl-B2 compared with earlier experiments. All results are presented in the same styles as in Fig. 2. Again, the lines for the GGA-PBESol represent also our very similar results obtained using the GGA-PBE- and OptB86b functionals, and the earlier calculations [40,55]. $C'(P)$ and $C_{44}(P)$ are highlighted by red and orange colors, respectively. Green crosses show radial-XRD data [62] where $A$-values at the highest pressures are outside of the selected range. See Fig. S5 in Supplemental Material for more details.

We also derived $G(P)$ of both NaCl phases (Fig. 3b) using the Hill approximation [63]. All previous and applied here DFT functionals provided lower $G(P)$ for NaCl-B1 and higher values for NaCl-B2. Our experimental $G(P)$ of NaCl-B2 agrees perfectly with the unique BLS data obtained on polycrystalline NaCl-B2 [6] (Fig. 3b). Apparently, a relatively weak elastic anisotropy of NaCl-B2 led to minor asymmetry BLS peaks from the polycrystalline samples and thus resulted in a reasonable $G(P)$ [20]. This agreement further supports our present TDBS results. Finally, our experimental $G(P)$ confirms the radial-XRD results [62] for NaCl-B1 while agreement for NaCl-B2 is rather qualitative due to a strong scattering of the radial-XRD data.

## IV. DISCISSION

### A. Discrepancies between experiments and DFT-calculations

In order to address reliability of modern DFT approaches, we compared existing experimental and theoretical results on pressure dependence of the principal shear moduli, $C_{44}$ and $C'$, of NaCl with those of three other compounds having similar bonding nature, namely CaO, MgO, and $(Mg_{1-x},Fe_x)O$. More precisely, we compared the dependences of $C_{44}$ and $C'$ on compression, the density ratio at high- and atmospheric pressure, $\rho/\rho_0$, up to the maximal value of ~1.5 (Fig. 4). Such comparison is meaningful because change of density, or of size of the unit cell, controls material properties including elastic moduli. One could notice that, in general, the shear moduli $C_{44}$ and $C'$ (as well as the anisotropy ratio $A$) of the four compounds change with $\rho/\rho_0$ in a similar manner: While $C'$ increases strongly with $\rho/\rho_0$, $C_{44}$ does not show a significant dependence. This similarity could be related to the same B1 crystal structure of the four compounds. Different values of the moduli and rate of change with compression, as well as the $\rho/\rho_0$ values at which $C_{44}$ and $C'$ cross we suppose to depend on composition and bonding strength of each particular compound. Analysis of these dependences permits also evaluation of capability of the used DFT approaches [40,55,64-70] to match experimental data and recognize perhaps any systematics.

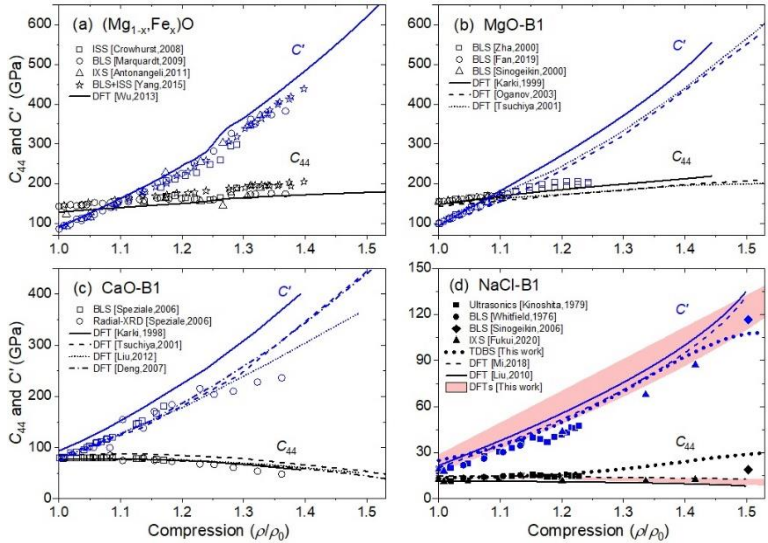

**FIG. 4**. Experimental and theoretical shear moduli $C_{44}$ (black symbols and lines) and $C'$ (blue symbols and lines) of B1 phases of (a) $(Mg_{1-x},Fe_x)O$ [28,29,32,70,71], (b) MgO [66,68,69,72-74], (c) CaO [64,65,67,69,75] as functions of $\rho/\rho_0$ compared with the present and earlier results for (d) NaCl [14-16,40,53,55]. Our experimental results for NaCl-B1 are shown by thick dotted curves and the theoretical

ones are summarized as light-red bands.

Analysis of the existing results for the four compounds suggests that all applied DFT approaches provided values systematically deviating from experimental ones for, at least, one of the two principal shear moduli, $C_{44}(P)$ or $C'(P)$, when $\rho(P)/\rho_0$ increases significantly (Fig. 4): They provide either systematically higher $C'(P)$ or systematically lower $C_{44}(P)$. In most cases, this leads to lower $A(P)$ except the DFT results where both $C'(P)$ and $C_{44}(P)$ are over- or underestimated. In the present work, this tendency was thoroughly elaborated for NaCl-B1 compressed to $\rho(P)/\rho_0 > 1.5$. However, this is not the case for MgO, $(Mg_{1-x},Fe_x)O$, and CaO: $C_{ij}(P)$ of MgO were measured using BLS to $\rho(P)/\rho_0 \sim 1.25$ only [74]. $C_{ij}(P)$ of several samples of $(Mg_{1-x},Fe_x)O$ with different Fe-contents were examined using various techniques to higher $\rho(P)/\rho_0$ but still well below 1.5. All measurements provided consistent results, e.g. similar dependences of $C'$ and $C_{44}$ on $\rho/\rho_0$ [28,29,32,71], but quite different from the dependence of $C'$ of MgO when the latter was compressed above $\rho(P)/\rho_0 = 1.1$ [74]. In contrast, there is only one recent DFT-calculation of $C_{ij}(P)$ of $(Mg_{1-x},Fe_x)O$ to very high $\rho(P)/\rho_0$ [70]. The situation is inverse for CaO: there is only one set of experimental $C_{ij}(P)$ measured using the classical BLS just to $\rho(P)/\rho_0 \sim 1.16$ [75]. In the same work, $C_{ij}(P)$ were derived, with significant uncertainties, from radial-XRD measurements on polycrystalline samples compressed to $\rho(P)/\rho_0 \sim 1.36$ [75]. Our present measurement on NaCl-B1 provided strong evidence that, upon compression, DFT approaches predict significantly lower $C_{44}(P)$ values when compared to the experiments, while predicted $C'(P)$ only slightly exceed the experimental ones. As a result, the theoretical $A(P)$ is underestimated (Fig. 3c). This tendency is also supported by recent measurements of $C_{ij}(P)$ of cubic solids exhibiting different bonding nature and $A > 1$, namely $H_2O$-ice [37] and solid argon [38]. The revealed systematic difference between DFT-calculations and experiments could also raise concerns by interpretation of seismological observations because $C_{ij}(P)$ of MgO and/or $(Mg_{1-x},Fe_x)O$ appear to control elastic anisotropy of rocks of the Earth's mantle and, thus, could provide access to the mantle texturing and convection [25,76,77].

## B. B1-B2 phase transition

Comparison of experimental pressure dependences $C_{44}(P)$ and $C'(P)$ of the above considered compounds permitted also proposing a mechanism of the B1-B2 phase transition and to predict pressures of the presently not verified transitions of MgO and $(Mg_{1-x},Fe_x)O$. in particular, it suggests that the B1-B2 phase transitions are initiated by a simple shear instability along one of the symmetry directions of the B1 structure manifesting itself in violation of the Born stability criterion $C_{44}(P) - P > 0$ [78]: According to our measurements, $C_{44}(P) - P$ of NaCl-B1 vanishes at $P \approx 30$ GPa (Fig. 5a), consistent with the experimental transition pressure $P_{tr}$, while the second Born stability criterion $C'(P) - P > 0$ (related to tetragonal shear) holds. Similarly, experimental $C_{44}(P) - P$ of CaO-B1 vanishes at $P \approx 58$ GPa perfectly matching $P_{tr} = 57.8$ GPa [75] while $C'(P) - P$ remains positive (Fig. 5b). All experimental data for $(Mg_{1-x},Fe_x)O$ suggest that only $C_{44}(P) - P$ becomes negative below 1 TPa (Fig. 5c) but its $P_{tr}$ is not determined yet: Using the most recent measurements of $C_{ij}(P)$ to the highest $\rho(P)/\rho_0$ [28], we predict $P_{tr}$ between 200 and 300 GPa applying a linear extrapolation of the last part of the experimental $C_{44}(P) - P$ dependence (Fig. 5c). The only relevant $C_{ij}(P)$ data for MgO-B1 [74] suggest an inverse order of violation of the Born criteria (Fig. 5d): $C'(P) - P$ should vanish as the first at a relatively low pressure of $P < 200$ GPa while $C_{44}(P)$ - $P$ well above 200 GPa (Fig. 5d). However, MgO-B1 is known to persist to $P = 228$ GPa at least [79]. Our extrapolation of the experimental $C_{44}(P)$ - $P$ suggests the transition pressure of $P_{tr} \sim 410$ GPa while the uncertainties cover a broad $P$-region including $P_{tr} \sim 600$ GPa proposed from laser-driven ramp-compression measurements [80].

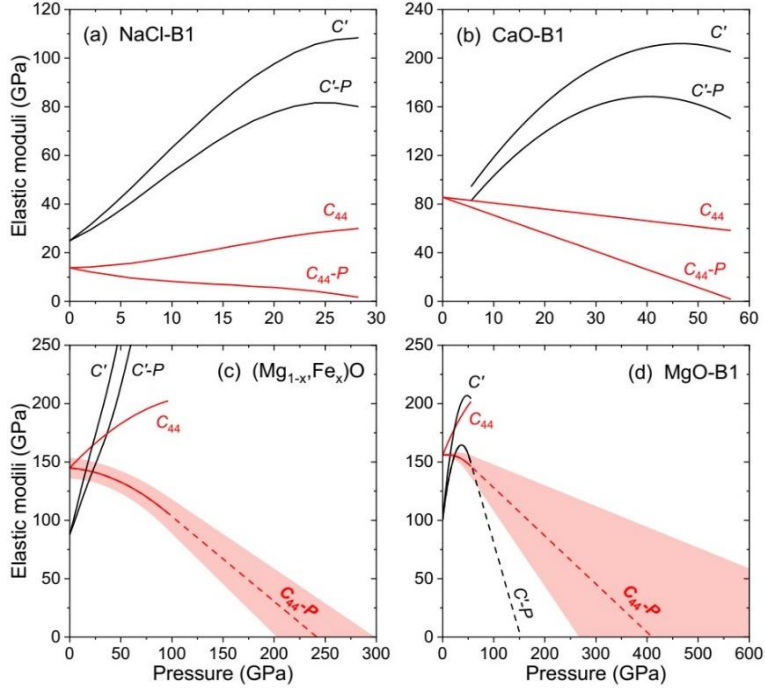

**FIG. 5.** Born stability criteria for B1 phases of (a) NaCl (this work), (b) CaO [75], (c) $(Mg_{1-x},Fe_x)O$ [28] and (d) MgO [74] derived from experiments. $C'(P)$ and $C'(P)$ - $P$ are shown by black lines, while $C_{44}(P)$ and $C_{44}(P)$ - $P$ by red lines. For $(Mg_{1-x},Fe_x)O$ and MgO, $C_{44}(P)$ - $P$ are shown by solid lines within experimental pressure regions and shaded bands represent the experimental uncertainties [28,74]. At higher pressures, dashed lines and shaded bands represent linear extrapolations of the last experimental data-points and of their uncertainties, respectively. For MgO, the experimental $C'(P)$ - $P$ and the extrapolations are presented in the same style.

Interestingly, the fact that violation of the criterion $C'(P)$ - $P > 0$ is not always a sign for instability of a phase can be recognized for other high-pressure cubic solids such as cubic $H_2O$ ice, stable in a broad pressure range between 2.2 and 80 GPa [37], and solid argon crystallising in the fcc structure at pressures from 1.3 to above 64 GPa [38]. Consequently, there are at least three compounds having cubic structures of different type (MgO-B1, ice VII and fcc solid argon) and persisting at pressures where the Born stability criterion $C'(P)$ - $P > 0$ is violated.

## V. CONCLUSIONS

Summarizing, our TDBS measurements on polycrystalline NaCl, compressed in a DAC to 41 GPa, yielded extremes of sound velocities in single crystals of both NaCl-B1 and NaCl-B2, $V_{L(100)}(P)$ and $V_{L(111)}(P)$. From these data, we derived the dependences $C_{44}(P)$, $C'(P)$, $G(P)$ and $A(P)$ and compared with our and previous DFT-calculations. We succeeded, therefore, to experimentally determine $C_{44}(P)$ and $C'(P)$ of NaCl-B2 not accessible for earlier-used experimental techniques because significantly big isolated single crystals are not accessible after the B1-B2 phase transition. This capability of the TDBS technique permits access to $C_{ij}(P)$ of any reconstructive cubic phase forming upon a reconstructive phase transition. We have found that NaCl-B1 exhibits a strong and growing with pressure elastic anisotropy, while that of NaCl-B2 is weaker and decreases with pressure. This supports the earlier proposal [6] to consider NaCl-B2 as an alternative pressure standard above 30 GPa. We used various advanced DFT-functionals, involving or not vdW forces, to predict $C_{44}(P)$ and $C'(P)$ of NaCl but none of them satisfactorily reproduced values of both B1- and B2 phase simultaneously. It would be premature to draw a definitive conclusion but the presented here results for NaCl, as well as the earlier results for other cubic solids [37,38,70,74,75], suggest a systematic failure of DFT approaches to predict principal shear moduli and elastic anisotropy of strongly compressed solids. Last but not least, our and earlier measurements indicate that violation of the Born stability criterion $C_{44}(P) - P > 0$ portends the B1-B2 transitions in NaCl, CaO, $(Mg_{1-x},Fe_x)O$ and, most probably, MgO.

## Acknowledgements

The authors acknowledge financial support from NSFC-41504070, NSFC-41874103, ZBYY-80922010601, CSC-201606955092 and the French National Research Agency, ANR (project I2T2M, No. ANR-18-CE42-0017). We thank E. Peronne for experimental assistance, V. E. Gusev, X. Liu and Y. Wu for discussions. We acknowledge access to the cluster MAGI of University Sorbonne Paris Nord and N. Greneche for the support.

Supplement to the manuscript
**"Single-crystal elastic moduli, anisotropy and the B1-B2 phase transition of NaCl at high pressures: Experiment vs. *ab-initio* calculations"**

Feng Xu,[1] Laurent Belliard,[2] Chenhui Li,[3,4] Philippe Djemia,[3] Loïc Becerra,[2] Haijun Huang,[1] Bernard Perrin[2] and Andreas Zerr[3]

[1]School of Science, Wuhan University of Technology, 430070 Wuhan, China

[2]Institut des NanoSciences de Paris, CNRS UMR 7588, Sorbonne Université, 75005 Paris, France

[3]Laboratoire des Sciences des Procédés et des Matériaux, CNRS UPR 3407, Université Sorbonne Paris Nord, Alliance Sorbonne-Paris-Cité, 93430 Villetaneuse, France

[4]School of Materials Science and Engineering, Beijing Institute of Technology, 100081 Beijing, China

In this document we provide:
- Detailed presentation of our DFT calculations, including
  - Methods,
  - Calculated enthalpies and EOSes,
  - Detailed presentation of calculated sound velocities and elastic parameters for all DFT functionals used in this work,
  - Calculation of the $C_{ij}(P)$ and development of the finite strain hydrostatic approach using the pure GGA-PBE functional.
- Synopsis of our experimental results

## [Detailed presentation of our DFT calculations]
### *Methods*
As concluded in the main text, we have recognized a limited ability of the used advanced density-functional-theory (DFT) approaches to predict one of the principal shear moduli $C_{44}$ or $C'$ or even both of them for strongly compressed B1 and B2 phases of NaCl. Below we describe in more details the efforts undertaken with the aim to verify that the selected computational parameters did not alter our finding.

Our *ab-initio* calculations were based on DFT as implemented in the Vienna Ab-initio Simulation Package (VASP) [1-3]. They were accomplished by direct calculations with VASP and with tools and functions interfaced to VASP within the Material Exploration and Design Analysis (MedeA®) software [4,5] which handles automatic execution of task sequences. Our DFT calculations on the geometry optimization, elastic and optical properties of both NaCl phases were performed as a function of $P$, using approaches based on PBE functionals (GGA-PBE, GGA-PBESol) [6,7], those with vdW corrections (GGA-PBE-TS-HI [8], GGA-PBE-D2, GGA-PBESol-D2 [9] and GGA-PBESol-D3 [10]), as well as the pure vdW functional (OptB86b) [11]. The hybrid functional HSE06 [12-14] was used to derive refractive index $n(P)$ of NaCl. We considered sodium with $sp$ valence electrons (Na_*sv* potential) and chlorine with seven valence electrons ($3s^23p^5$, Cl_h hard potential). The electron-ion interactions were described by the projector augmented wave method (PAW) [15] with a plane wave energy cutoff of 500 eV and 1200 eV for the HSE06 and PBE-GGA potentials, respectively, for the cell optimization. The energy convergence criterion for ionic relaxations was set to $10^{-3}$ meV/atom. The damped molecular dynamics algorithm and reciprocal space projection operators were used. The Monkhorst-Pack scheme [16] was applied to construct 5×5×5 and 11×11×11 $k$-meshes of the Brillouin zone for self-consistent and non-local exchange calculations, and higher $k$-meshes 40×40×40 and 30×30×30 for convergence tests using the GGA-PBE, the last one being retained. The $k$-meshes for the density of states and optical spectra were two times finer for non-local exchange, namely 10×10×10 and 22×22×22. The $k$-meshes were forced to be centered on the $\Gamma$ point of the Brillouin zone and a Gaussian smearing with the width of 0.2 eV was employed for relaxation of the examined crystal structures. The number of bands (NBANDS parameter) was set to 30. Frequency-dependent optical properties (here the refractive index $n$ at 1.55 eV, corresponding to the light wavelength of $\lambda$= 800 nm) were calculated using the VASP-TAG (LOPTICS=.TRUE.) and the complex shift, used to smoothen the real part of the dielectric function, was set to 0.1. A detailed description of the method can be found elsewhere [17]. Single-crystal elastic moduli $C_{ij}(P)$ were calculated applying the strain-stress method with moderate strains of 0.001-0.004 and strain matrixes proposed in Ref. [18]. The calculations were further extended to the strain-energy method [18] and much larger strains of [-0.08, 0.08] in order to calculate not only $C_{ij}(P)$ but also the third-order-elastic constants $C_{ijk}$, which were needed by development of the so-called hydrostatic model with finite strains.

### *Calculated enthalpies and EOSes*

Differences of the enthalpies ($F = E + PV$) of NaCl-B1 and NaCl-B2 at $T$=0 K as functions of hydrostatic pressure $P$, were calculated using several DFT functionals (Fig. S1). Here, $E$ is the total energy and $V$ volume of the unit cell. The phase transition pressure $P_{tr}$, from NaCl-B1 to NaCl-B2, was found to vary between 14 GPa and 26 GPa (Fig. S1). The DFT functionals without consideration of vdW forces (GGA-PBE and GGA-PBESol) provided $P_{tr}$~26 GPa and ~22 GPa, respectively, they are the closest to the experimentally observed $P_{tr}$~30 GPa. The pure vdW functional OptB86b provided $P_{tr}$~22 GPa while the vdW corrected functionals (GGA-PBE-D2 and GGA-PBE-TS-HI) provided even lower $P_{tr}$.

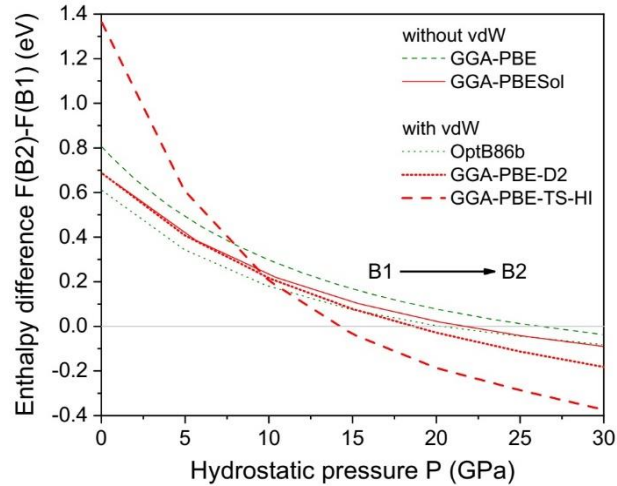

**Fig. S1.** Differences of enthalpies $F = E+PV$ of the B1 and B2 phases of NaCl at $T$=0 K, calculated using different functionals (GGA-PBE, GGA-PBESol, OptB86b, GGA-PBE-D2 and GGA-PBE-TS-HI), suggest transition pressures between 14 GPa and 26 GPa.

Earlier measured and calculated $B(P)$, as well as those obtained from our DFT-calculations using seven different functionals are shown in Fig. S2. Earlier shock-compression data of Fritz et al. [19] deviate from the Decker's EOS at $P>15$ GPa, and reach a maximal difference of ~7%. The deviation can further grow to ≥17% if the experimental uncertainties are taken into account. Our calculations using the GGA-PBESol, GGA-PBE and OptB86b functionals produced similar results agreeing well with the Decker's EOS for the B1 phase and with the EOS of Fei et al. for the B2 phase [20]. Earlier calculations by Liu et al. [21] and Mi et al. [22] also provided similar $B(P)$. Other DFAs applying the vdW corrected functionals (GGA-PBE-TS-HI, GGA-PBE-D2, GGA-PBESol-D2 and GGA-PBESol-D3) provided less consistent data deviating from both the Decker's EOS and the shock-compression data of Fritz et al. [19].

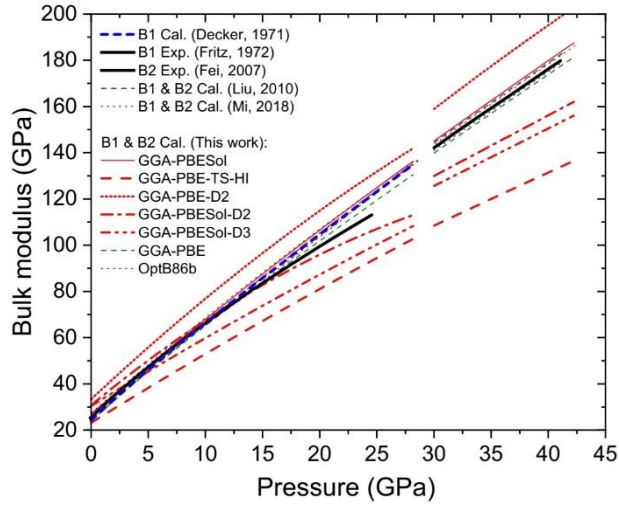

**Fig. S2.** Bulk moduli, $B(P)$, of the B1- and B2 phases of NaCl. That from the Decker's EOS of NaCl-B1 is presented by dashed blue line. $B(P)$ of NaCl-B1 from the shock compression experiments of Fritz et al. [19], as well as that from the EOS of NaCl-B2 obtained via static compression by Fei et al. [20] are shown by solid black lines. Recent theoretical $B(P)$ of both phases based on DFAs are presented by thin violet dashed and dotted lines [21,22]. Our theoretical results are shown as follows: GGA-PBESol, GGA-PBE-TS-HI, GGA-PBE-D2, GGA-PBESol-D2 and GGA-PBESol-D3 functionals are represented by red solid(thin), dashed, dotted, dash-dotted and dash-dot-dotted lines, respectively; while GGA-PBE, and OptB86b by thin green dashed and dotted lines, respectively.

***Detailed presentation of calculated sound velocities and elastic parameters for all DFT functionals used in this work***

In Figs. 2 and 3 of the main text, we did not show separately our DFT-results obtained using the GGA-PBE- and OptB86b functionals and the earlier predictions [21,22], because they are nearly identical to the presented values obtained using the GGA-PBESol functional. In particular, the latter results are typically between those obtained using the GGA-PBE and OptB86b functionals. In Fig. S3, we show all available experimental and theoretical values of $V_{L(100)}(P)$ and $V_{L(111)}(P)$. Compared to Fig. 2 of the main text, we added in Fig. 3 results of all our and earlier DFAs but did not show the $V_{L(avg)}$ values for the sake of clarity. Similarly, we show all available results in Figs. S4 and S5, including the experimental radial-XRD data reported in the thesis work of Mi [23].

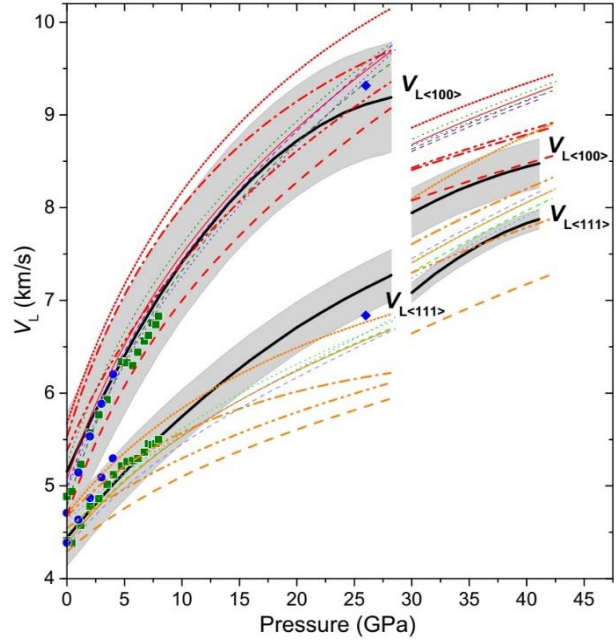

**Fig. S3.** Longitudinal sound velocities $V_{L(100)}$ and $V_{L(111)}$ of compressed NaCl-B1 and NaCl-B2. Our experimental and theoretical results are shown in the same style as in Fig. S2. Earlier measurements on single crystals using BLS (solid blue circles and diamonds) and ultrasonics (solid green squares) agree well with our experiments. For better readability, all theoretical (our and previous) $V_{L(111)}(P)$ dependences are shown using lighter colors when compared to those for $V_{L(100)}(P)$.

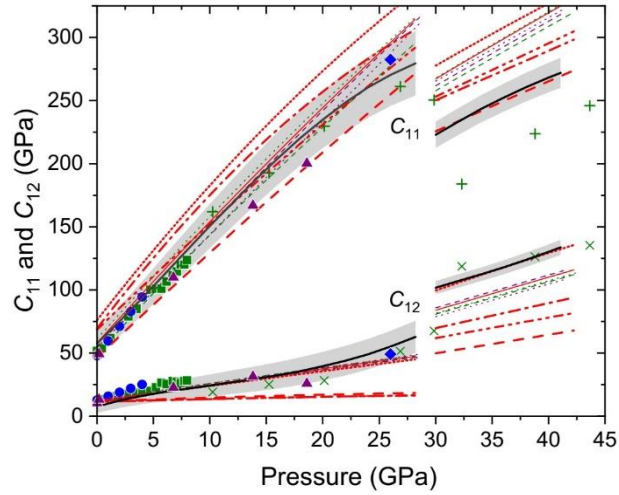

**Fig. S4.** Elastic moduli $C_{11}(P)$ and $C_{12}(P)$ of NaCl-B1 and NaCl-B2. Experimental and theoretical results of this work and the earlier reported ones are shown in the same style as in Figs. S2 and S3. Radial-XRD data reported by Mi [23] are presented by + signs.

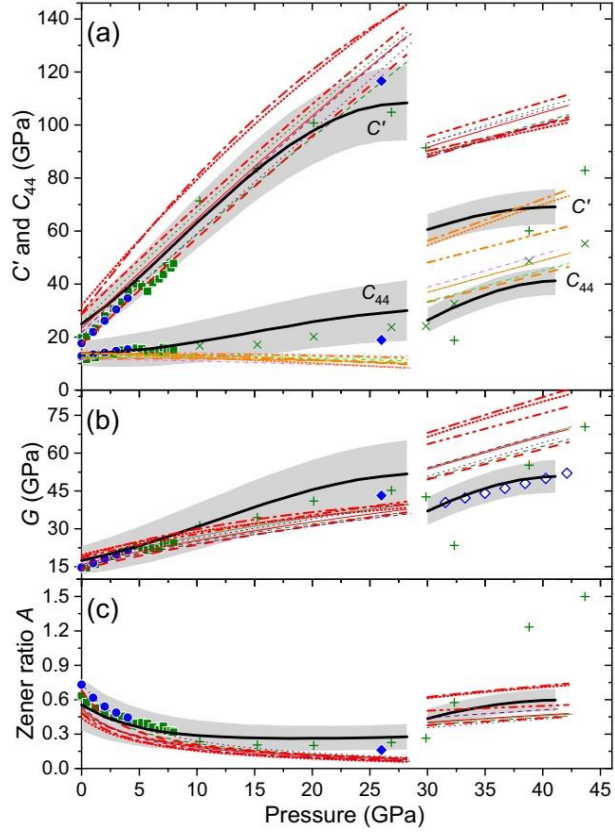

**Fig. S5.** Elastic moduli $C'(P)$, $C_{44}(P)$, $G(P)$ and Zener anisotropy ratio $A(P)$ of NaCl-B1 and NaCl-B2. Experimental and theoretical results of this work and the earlier reported ones are shown in the same style as in Figs. S2 and S3. In order to better distinguish the lines in (a), all theoretical (our and earlier) $C_{44}(P)$ values are drawn using lighter colors when compared with those for $C'(P)$. For radial-XRD data reported by Mi [23], $C_{44}(P)$ and $C'(P)$ are presented by + and × signs.

### *Calculation of the $C_{ij}(P)$ and development of the finite strain hydrostatic approach using the pure GGA-PBE functional*

In the case of materials under hydrostatic compression and finite strain, it is convenient to describe their elastic behavior using pressure dependences of elastic moduli $C_{ij}(P)$ and of bulk modulus $B(P)$. In most cases, it is sufficient to consider their linear dependence on pressure $P$:

$$\boldsymbol{C_{11}(P) \approx c_{11} + c'_{11}(P - P_0)}$$

$$\boldsymbol{C_{12}(P) \approx c_{12} + c'_{12}(P - P_0)}$$

$$\boldsymbol{C_{44}(P) \approx c_{44} + c'_{44}(P - P_0)}$$

$$B(P) \approx \frac{[C_{11}(P) + 2C_{12}(P)]}{3} \qquad \text{Eqs. (S1)}$$

where $c_{ij}$ are the second-order elastic constants (SOECs) and $c'_{ij}$ their pressure derivatives at a starting pressure $P_0$. For NaCl-B1, the starting pressure is $P_0$=0 GPa while for NaCl-B2 we selected $P_0$=32 GPa. This approximation, which is much less computationally demanding, we denote in the following as the hydrostatic approach. Its results are compared below with our direct DFT calculations of $C_{ij}(P)$ presented above. In general, the pressure derivatives $c'_{ij}$ are determined through the SOECs and the third-order elastic constants $C_{ijk}$ (TOECs) at $P=P_0$, as defined by Birch [24]:

$$c'_{11} = \frac{dC_{11}}{dP} = -\frac{(2C_{112} + C_{111} + 2c_{12} + 2c_{11})}{(2c_{12} + c_{11})}$$

$$c'_{12} = \frac{dC_{12}}{dP} = -\frac{(2C_{112} + C_{123} - c_{12} - c_{11})}{(2c_{12} + c_{11})}$$

$$c'_{44} = \frac{dC_{44}}{dP} = -\frac{(2C_{155} + C_{144} + c_{44} + 2c_{12} + c_{11})}{(2c_{12} + c_{11})} \qquad \textbf{Eqs. (S2)}$$

We performed the DFT calculations of the SOECs and TOECs in pressure steps of 5 GPa applying two methods as exposed in detail in Refs. [18,25,26]. Namely, we used (i) the strain-energy method ($\boldsymbol{\eta}$-$E/V_0$) where $V_0$ is the volume of the unstrained crystal ($\boldsymbol{\eta = 0}$) and $E$ the total energy, and (ii) the strain-stress ($\boldsymbol{\eta}$-$t$) method. Only SOECs from the latter method are presented in the Main text for the investigated pressure range. In both methods, six different strain matrixes ($\boldsymbol{\eta_{A-F}}$) were applied varying the strain $\eta$ between [-0.08, 0.08] and assuming the sample material is hyper-elastic. Three SOECs and six TOECs were obtained from the fitting of the curves ($\boldsymbol{\eta}$-$E/V_0$) and ($\boldsymbol{\eta}$-$t$). In Fig. S6, we show, for illustration, curves obtained at zero pressure, in case of the strain matrix $\boldsymbol{\eta_A}(\eta, 0,0,0,0,0)$, energy per unit volume $\frac{E(\eta_A)}{V_0} = \frac{1}{6}C_{111}\eta^3 + \frac{1}{2}c_{11}\eta^2$ and stress $t_1(\eta_A) = \frac{1}{2}C_{111}\eta^2 + c_{11}\eta$. For NaCl-B1, the calculations were performed in the pressure range from 1 atm. to 30 GPa while for NaCl-B2 the covered pressure range was between 32 GPa and 50 GPa (Fig. S7). For the above mentioned hydrostatic approach, only the DFT calculations at $P_0$=1 bar, for NaCl-B1, and at $P_0$=32 GPa, for NaCl-B2, were necessary. All values needed in the hydrostatic approach are summarized in Table S1 and compared with earlier theoretical and experimental data. It has to be mentioned that $C_{111}$ (for which we obtained -692±20 GPa at zero pressure) is quite sensitive to the applied strain range which, however, is also the case for other $C_{ijk}$. Such sensitivity can lead to higher $C_{ijk}$ values if compressive strain is favored at expense of the tensile part: For example, we obtained $C_{111}$ = -830 GPa for the strain range [-0.08, 0.05], which is much closer to the earlier reported experimental data.

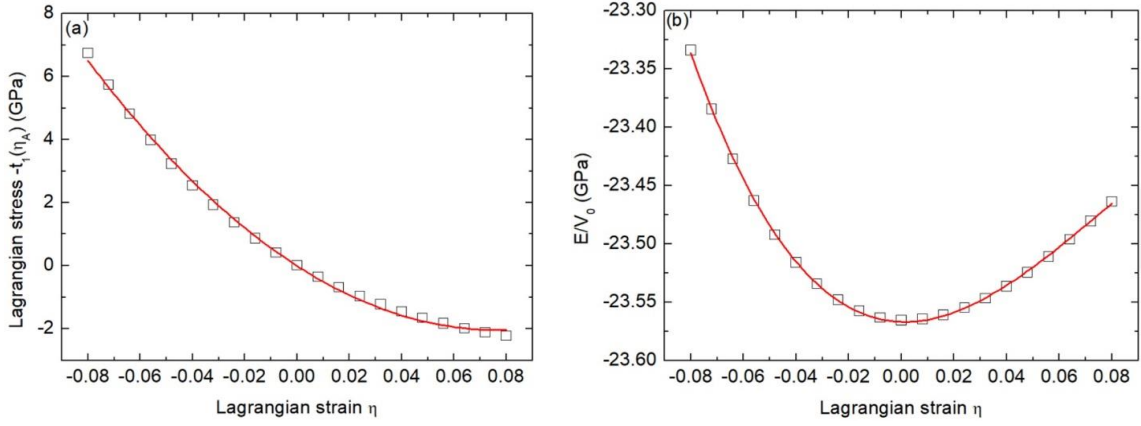

**Fig. S6.** (a) Lagrangian stress *vs*. Lagrangian strain, $-t_1(\mathbf{\eta}_A)$, at zero pressure. Symbols denote results of our DFT calculations using the PBE-GGA functional; red solid line represents the second-order polynomial fit ($350\eta^2$-$53\eta$) to these points. (b) $E/V_0$ *vs*. Lagrangian strain, $E^*(\mathbf{\eta}_A)$, at zero pressure. Symbols denote results of our DFT calculations using the PBE-GGA functional; solid line represents third order polynomial fit ($-115\eta^3$+$25\eta^2$) to these data-points. Fitting of these strain-energy and strain-stress curves, and of all other curves of this type were performed with a step of 0.008. Definitions of Lagrangian stress $t_1(\mathbf{\eta}_A)$, strain-energy relation $E^*(\mathbf{\eta}_A)$ and strain matrix $\mathbf{\eta}_A$ can be found elsewhere [25].

In Fig. S7, we compare $C_{ij}(P)$ calculated using both models, strain-stress and strain-energy, with those obtained from the hydrostatic approach where linear pressure dependences of the moduli $C_{ij}$ are assumed (see Eqs. S1). The models could reproduce the experimental dependences reported in the Main text only partially: There is a significant disagreement for the $C_{44}(P)$ of the B1 phase, especially for the hydrostatic approach, and for the $C_{11}(P)$ (and consequently $C_{12}(P)$) of the B2 phase. Thus, our TDBS measurements provided experimental tests of the used here computational approaches as is also the case for MgO-B1 where the DFAs predicted correctly $C_{11}(P)$ and $C_{12}(P)$ but underestimated $C_{44}(P)$ [27,28]. Comparison of results of these labor-intensive DFT calculations, based on the strain-stress and strain-energy methods with very large strains from -0.08 to 0.08, indicates that, for the examined material, the approach presented in the Main text (with a low strain $\eta$ from -0.004 to 0.004) deviates less from the experimental data. However, the main drawback, namely systematic underestimation of $C_{44}(P)$ of the B1 phase and systematic overestimation of $C'(P)$ of the B2 phase (Fig. S5) could not be resolved. The hydrostatic approach, considered as a simple approximation to describe elasticity of the NaCl phases at high pressures, is even less accurate.

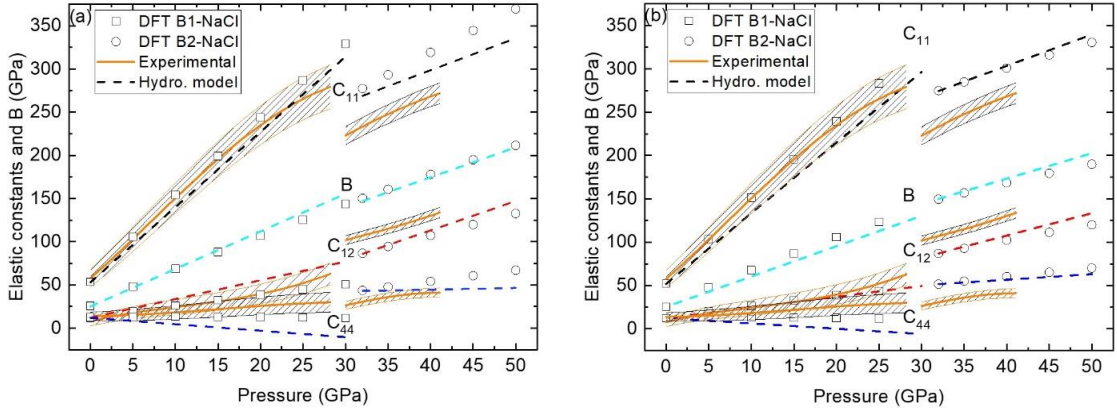

**Fig. S7.** $C_{ij}(P)$ from our DFT calculations using the PBE-GGA functional (symbols): (a) strain-stress method and (b) strain-energy method. Dashed lines represent linear dependences of $C_{ij}$ obtained using the Eqs. S1 and the finite-strain hydrostatic approach. Experimental results are represented by orange lines and the experimental uncertainties are shown by shadowed bands.

**Table S1:** Calculated SOECs $c_{ij}$, their pressure derivatives, $c'_{ij}$, and the six TOECs, $C_{ijk}$, of cubic NaCl-B1 and NaCl-B2 obtained using the hydrostatic approach. We show results obtained using both methods, strain-energy and strain-stress, and compare with earlier reported experimental data. $C_{ijk}$ and $c_{ij}$ values are given in GPa, mass density in g/cm³ and the unit-cell volume in Å³.

| NaCl | Present theoretical results at $T$=0K | | Experimental values (room temperature) | |
|---|---|---|---|---|
| | Strain-energy | Strain-stress | Present | Others |
| B1 ($P_0$=0 GPa) | | | | |
| Mass density/$V_0$ | 2.1060/46.08 | 2.1060/46.08 | | |
| | | | | |
| $c_{11}$ | 52±0.3 | 53±0.5 | 57 | 49.8±0.5[b], 51.6±0.5[c], 48.2±1.4[d] |
| $c_{12}$ | 12±0.4 | 12±0.5 | 8 | 13.0±0.2[b], 12.2±0.12[c], 12.8±0.8[d] |
| $c_{44}$ | 12±0.1 | 12±0.3 | 14 | 12.8±0.1[b], 13.6±0.2[c], 12.7±3.7[d] |
| B | 26±0.7 | 25±1 | 24.3 | 37.9[b], 25.3[c], 24.6±0.7[d] |
| | | | | |
| $c'_{11}$ | 8.15(10.30[a]) | 8.71(9.92[a]) | 8.83[a] | 9.15[b], 8.72[c], 11.62±0.7[d] |
| $c'_{12}$ | 1.24(0.47[a]) | 2.17(1.50[a]) | 2.37[a] | 0.076[b], 1.22[c], 3.05±0.37[d] |
| $c'_{44}$ | -0.61(0.63[a]) | -0.75(0.104[a]) | 0.147[a] | 0.14[b], 0.31[c], 0.759±0.046[d] |
| $B'$ | 3.55 (3.88[a]) | 4.35(4.30[a]) | 4.52[a] | 3.1[b], 3.54[c], 5.9±0.4[d], 5.6[e] |
| | | | | |
| $C_{111}$ | -692±20 | -700±24 | | **-823±2[b], -830±80**[f] |
| $C_{112}$ | -28±8 | -51±10 | | 2±5[b] |
| $C_{144}$ | 22±2 | 15±5 | | 23±3[b] |
| $C_{155}$ | -32±3 | -23±3 | | -61±3[b] |
| $C_{123}$ | 26±35 | 3±3 | | 53±7[b] |
| $C_{456}$ | 23±0.1 | 58±2 | | 20±1[b] |
| B2 ($P_0$=32 GPa) | | | | |
| Mass density/$V_0$ | 3.5060/ 27.68 | 3.5060/ 27.68 | | |
| $c_{11}$ | 275±3 | 269±3 | 233 | |
| $c_{12}$ | 87±2 | 86±2 | 107 | |
| $c_{44}$ | 51±1 | 43±1 | 30 | |
| B | 150±2 | 147±2 | 149 | |
| | | | | |
| $c'_{11}$ | 3.62(3.44[a]) | 3.7(5.14[a]) | 5.18[a] | |
| $c'_{12}$ | 2.58(1.23[a]) | 3.34(2.57[a]) | 2.64[a] | |
| $c'_{44}$ | 0.64(1.55[a]) | 0.19(1.16[a]) | 2.15[a] | |
| $B'$ | 2.93(1.96[a]) | 3.46(3.43[a]) | 3.49[a] | |
| | | | | |
| $C_{111}$ | -1670±50 | -1675±50 | | |
| $C_{112}$ | -191±7 | -356±10 | | |
| $C_{144}$ | -301±20 | -317±20 | | |
| $C_{155}$ | -199±18 | -131±16 | | |
| $C_{123}$ | -282±20 | -424±22 | | |
| $C_{456}$ | -367±3 | -667±6 | | |

[a] Initial slope obtained from polynomial fits to the points calculated using the GGA-PBE functional and from our experimental $C_{ij}(P)$ in the 0-5 GPa range, for NaCl-B1, and in the range 32-37 GPa, for NaCl-B2, [b] Ref. [29] and references herein, [c] Ref. [30], [d] Ref. [31], [e] Ref. [32], [f] Ref. [33].

## [Synopsis of our experimental results]

Tables S2 (for NaCl-B1) and S3 (for NaCl-B2) summarize our experimental data, as well as the theoretical refractive indexes used in this work, in form of polynomials of the type $f(P) = Intercept + I_1P + I_2P^2 + I_3P^3 + I_4P^4$ obtained using the least-square fits. The sound velocities $V_{L(100)}$ and $V_{L(111)}$ are given in km/s, elastic constants $C_{11}$, $C_{12}$, $C_{44}$ and averaged modulus $G$ (using Hill approximation) in GPa. Refractive index $n$ and Zener anisotropy ratio $A$ are dimensionless.

**Table S2** (for NaCl-B1, from 1 atm. to 28.2 GPa,)

| | Intercept | $I_1$ | $I_2$ | $I_3$ | $I_4$ |
|---|---|---|---|---|---|
| $n$ | 1.575 | 0.007 | -7.03 x$10^{-5}$ | | |
| $V_{L<100>}$ | 5.18 | 0.27 | -47 | | |
| $V_{L<111>}$ | 4.46 | 0.15 | -19.1 | | |
| $C_{11}$ | 57.26 | 8.85 | 9.15x$10^{-2}$ | -47.7 | |
| $C_{12}$ | 7.99 | 2.31 | -97.6 | 2.93x$10^{-3}$ | |
| $C_{44}$ | 13.79 | 8.91x$10^{-2}$ | 4.49x$10^{-2}$ | 9.86x$10^{-4}$ | |
| $G$ | 17.5 | 0.93 | 5.55x$10^{-2}$ | -19.2 | |
| $A$ | 0.551 | -58.6 | 4.29x$10^{-3}$ | 1.47x$10^{-4}$ | 1.92x$10^{-6}$ |

**Table S3** (for NaCl-B2, from 30 to 41.1 GPa)

| | Intercept | $I_1$ | $I_2$ |
|---|---|---|---|
| $n$ | 1.663 | 0.004 | -28.5 |
| $V_{L<100>}$ | 4.35 | 0.17 | 1.75x$10^{-3}$ |
| $V_{L<111>}$ | 1.21 | 0.29 | -33.3 |
| $C_{11}$ | -22.76 | 10.93 | -0.09 |
| $C_{12}$ | 63.32 | 0.13 | $3.85 \times 10^{-2}$ |
| $C_{44}$ | -134.55 | 8.28 | -99.3 |
| $G$ | -117.75 | 8.01 | -97 |
| $A$ | -1.39 | $9.46 \times 10^{-2}$ | $1.13 \times 10^{-3}$ |

**[References]**

(1999).